\documentclass[]{mosi}

\usepackage{helvet}

\usepackage{amsmath} 
\usepackage{natbib}
\usepackage{graphicx}
\usepackage{subcaption} 

\usepackage[toc,page,header]{appendix}
\usepackage[utf8]{inputenc} % allow utf-8 input
\usepackage[T1]{fontenc}    % use 8-bit T1 fonts
\usepackage{hyperref}       % hyperlinks
\usepackage{url}            % simple URL typesetting
\usepackage{booktabs}       % professional-quality tables
\usepackage{lmodern}        % scalable Latin Modern fonts
\usepackage{amsfonts}       % blackboard math symbols
\usepackage{nicefrac}       % compact symbols for 1/2, etc.
\usepackage{microtype}      % microtypography
\usepackage{wrapfig}

\usepackage{amssymb}        % special symbols
\usepackage{titletoc}
\usepackage{minitoc}

\usepackage{array}
\usepackage{etoolbox}

\definecolor{lightblue}{RGB}{200, 230, 255}  
\definecolor{headerblue}{RGB}{150, 200, 255} 

\usepackage{pgfplots}
\usepackage{xcolor}
\usepackage{float}
\usepackage{comment}
\usepackage{multirow}
\usepackage{makecell}
\usepackage{siunitx}
\usepackage{tikz}
\usepackage{pgf-pie}
\usepackage[export]{adjustbox}

\usepackage{ragged2e}
\usepackage{tabularx}
\usepackage{caption}
\usepackage{enumitem}
\usepackage{pifont}
\usepackage[hang,flushmargin]{footmisc}

\usepackage{tcolorbox}
\tcbuselibrary{breakable}
\tcbuselibrary{skins}

\usepackage{tabularx}
\usepackage{listings}

\usepackage{threeparttable}
\usepackage{setspace}
\usepackage[scheme=plain]{ctex} % needed for Chinese text in appendix prompts

\definecolor{MossCyan}{HTML}{82D9FF} 
\definecolor{MossBlue}{HTML}{82B1FF}

\definecolor{ForestGreen}{RGB}{34, 139, 34}
\definecolor{Red}{RGB}{255, 0, 0}

\definecolor{tickG}{rgb}{0.1, 0.588, 0.1}
\definecolor{crossR}{rgb}{0.588, 0.1, 0.1}

\definecolor{frenchblue}{rgb}{0.0, 0.45, 0.73}
\definecolor{babyblue}{rgb}{0.54, 0.81, 0.94}
\definecolor{classicrose}{rgb}{0.98, 0.8, 0.91}
\definecolor{beige}{rgb}{0.96, 0.96, 0.86}
\definecolor{forestgreen}{HTML}{2e7d43}

\definecolor{blue1}{HTML}{91BBE6}
\definecolor{blue2}{HTML}{3F90E0}
\definecolor{blue3}{HTML}{316FAD}

\definecolor{color1}{HTML}{FF9999}
\definecolor{color2}{HTML}{FF6666}
\definecolor{color3}{HTML}{FF3333}
\definecolor{color4}{HTML}{E60000}
\definecolor{color5}{HTML}{B30000}
\definecolor{color6}{HTML}{8CD98C}
\definecolor{color7}{HTML}{53c653}
\definecolor{color8}{HTML}{00B050}
\definecolor{color9}{HTML}{2d862d}
\definecolor{color10}{HTML}{206020}
\definecolor{color11}{HTML}{cca300}

\newtcolorbox{promptbox}[2][]{
    colback=white,
    coltext=black,
    arc=3mm,
    boxrule=0.5pt,
    colframe=black!60!white,
    title={#2},
    colbacktitle=black,
    coltitle=white,
    fonttitle=\bfseries,
    top=8pt,
    bottom=8pt,
    left=10pt,
    right=10pt,
    breakable,
    before upper={%
        \linespread{1}\selectfont
        \setlength{\parskip}{1ex plus 0.2ex minus 0.2ex}%
        \setlength{\parindent}{0pt}%
    },
    #1
}

\title{MOSS Transcribe Diarize Technical Report}

\author{MOSI.AI\textsuperscript{*}}

\abstract{
\begin{abstract}
'Speaker-Attributed, Time-Stamped Transcription (SATS) aims to transcribe what is said and to precisely determine the timing of each speaker, which is particularly valuable for meeting transcription. Existing SATS systems rarely adopt an end-to-end formulation and are further constrained by limited context windows, weak long-range speaker memory, and the inability to output timestamps. To address these limitations, we present \textbf{MOSS Transcribe Diarize}, a unified multimodal large language model that jointly performs Speaker-Attributed, Time-Stamped Transcription in an end-to-end paradigm. Trained on extensive real wild data and equipped with a 128k context window for up to 90-minute inputs, MOSS Transcribe Diarize scales well and generalizes robustly. Across comprehensive evaluations, it outperforms state-of-the-art commercial systems on multiple public and in-house benchmarks.
\end{abstract}
}

\checkdata[Github]{\url{https://github.com/OpenMOSS/MOSS-Transcribe-Diarize}}
\checkdata[HuggingFace]{\url{https://huggingface.co/OpenMOSS-Team/MOSS-Transcribe-Diarize}}

\begin{document}
\maketitle
\begingroup
\renewcommand{\thefootnote}{\fnsymbol{footnote}}
\setcounter{footnote}{1}
\footnotetext{Full contributors can be found in the Contributors section.}
\endgroup

% ===== Original meta lines (content unchanged) =====
% \begin{spacing}{1.0}
% {\small \noindent \textbf{Date:} December 31, 2025 \par}
% {\small \noindent \textbf{Demo:} \url{https://openmoss.github.io/MOSS-Transcribe-Diarize-demo/} \par}
% {\small \noindent \textbf{Huggingface Space:} \url{https://huggingface.co/spaces/OpenMOSS-Team/MOSS-transcribe-diarize} \par}
% \end{spacing}

% ===== Original content starts here (sections/tables/figures unchanged) =====

\begin{figure}[htbp]
    \centering
    \includegraphics[width=0.8\textwidth]{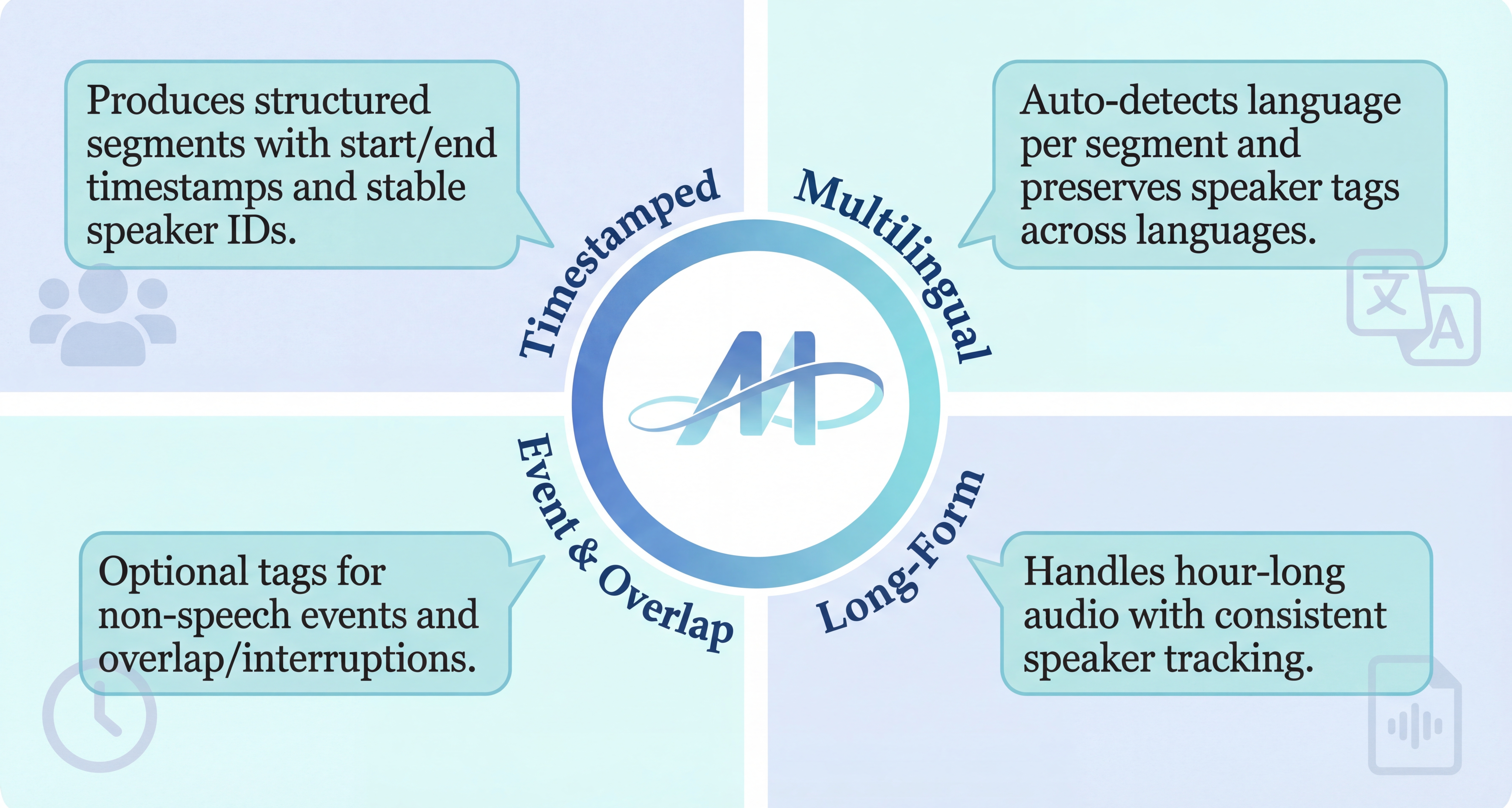}
    \caption{\textbf{Overview of key capabilities of MOSS Transcribe Diarize}}
    \label{fig:model_structure}
\end{figure}

\section{Introduction}
\noindent Accurate transcripts of multi-speaker conversations are foundational to a wide range of applications, from meeting assistants and call-center analytics to assistive technologies and legal discovery \citep{carletta2005ami,janin2003icsi,chen2020libricss}. In these settings, who said what, and when is as important as what was said: users need speaker-attributed, time-stamped transcripts that preserve turn structure, overlaps, and long-range references across a discussion that may span tens of minutes. We refer to this task as \textbf{Speaker-Attributed, Time-Stamped Transcription (SATS)}. Despite its practical importance, SATS is typically solved today by stitching together multiple components—automatic speech recognition (ASR) (e.g., Whisper \citep{radford2023whisper}) and speaker diarization (e.g., Pyannote \citep{bredin2020pyannote} or x-vector clustering \citep{snyder2018xvector}), each trained with different objectives and latencies, and often tuned on different datasets. Such modular pipelines are brittle: errors cascade across stages, global context is hard to leverage consistently, and end users must accept trade-offs between attribution accuracy, temporal precision, and throughput \citep{wang2024diarizationlm}.

\noindent Recent advances in large language models (LLMs) and multimodal large language models (MLLMs) \citep{chu2023qwenaudio,tang2023salmonn,zhang2023speechgpt} suggest a path toward unified solutions that jointly model audio and text. However, most existing MLLMs are developed and evaluated mainly on single-speaker speech, achieving strong ASR performance but falling short of our SATS setting, which requires jointly recognizing content, attributing it to speakers, and providing time-stamped speaker turns in multi-speaker conversations.

\noindent
A practical compromise between fully modular pipelines and fully end-to-end SATS is a semi-cascaded (or hybrid) scheme: a strong ASR and an acoustic diarization front-end are still used to produce candidate words and speaker traces, while an LLM-style model is introduced as a global reconciliation layer to resolve speaker permutations, repair boundary inconsistencies around turns/overlaps, and improve the readability and consistency of the final speaker-attributed transcript. For example, DiarizationLM post-processes the independent outputs from ASR and diarization systems with an LLM to refine speaker-attributed transcriptions, but it remains non end-to-end and thus inherits the classic error-propagation and mismatch issues of cascaded designs \citep{wang2024diarizationlm}. 

\noindent
Motivated by the brittleness of modular ASR--diarization pipelines, recent work has begun to unify recognition and speaker attribution within a single multimodal framework, so that lexical modeling and speaker attribution can be learned jointly under shared context. This line of research moves SATS closer to an end-to-end formulation and reduces cross-module mismatch.
Sortformer~\cite{park2025sortformer} explores joint modeling with a permutation-invariant objective (Sort Loss) to better align speaker identity with lexical tokens, but its training remains two-stage: it first trains a diarization-only model and then freezes it to provide speaker traces as inputs for training an ASR model. As a result, it is not a truly end-to-end SATS formulation and can still suffer from cross-stage mismatch. SpeakerLM~\cite{yin2025speakerlm} is closer to a unified architecture by integrating speaker-aware modeling into a single MLLM, demonstrating improved speaker attribution without an explicit modular diarization stage; however, in their reported settings these models are still limited to relatively short audio contexts (on the order of 50--90\,s) and small speaker sets (e.g., up to 4 speakers), which restricts scalability to meeting-style, long-form conversations. Moreover, these approaches do not natively output explicit time-stamped speaker segments (i.e., ``who spoke when'') at the segment level, which is important for meeting-style SATS.

\noindent
To tackle long-form scalability within an MLLM-style framework, \citet{shi2025trainshortinferlong} propose JEDIS-LLM, an end-to-end Speech-LLM for joint ASR and diarization that is trained only on short clips ($\le$20\,s) yet supports chunk-wise, streamable inference on long-form audio via a Speaker Prompt Cache (SPC) with on-the-fly updates; SPC also enables integrating pre-enrolled speaker profiles commonly used in meeting transcription. While these hybrid/streaming designs substantially improve long-audio scalability under latency constraints, they still rely on chunk-wise processing and additional mechanisms (e.g., cache management and segmentation/alignment \citep{mcauliffe2017mfa,bain2023whisperx}) to maintain global speaker consistency. This further motivates our long-context, single-pass SATS formulation that avoids chunk boundaries and natively emits timestamped speaker turns.

\noindent 
These trends underscore that most existing SATS systems still fall short of a truly end-to-end formulation. Despite promising progress, current approaches are often constrained in three key aspects. First, limited context windows force long recordings to be processed in short chunks, which disrupts discourse continuity, weakens coreference resolution and speaker consistency, and introduces boundary artifacts for timestamping \citep{shi2025trainshortinferlong,bain2023whisperx}. Second, long-range speaker memory remains fragile: speaker identities may drift over dozens of turns, especially under similar voices or varying acoustic conditions \citep{snyder2018xvector,fujita2019eend}. Third, many architectures cannot natively produce segment-level timestamps with the granularity needed for retrieval and skimmability, and thus rely on external alignment components that re-introduce modular error propagation \citep{mcauliffe2017mfa,bain2023whisperx}.

\noindent
In this paper we address these limitations with \textbf{MOSS Transcribe Diarize}, a unified multimodal large language model designed to perform SATS in a single, end-to-end pass. MOSS Transcribe Diarize jointly recognizes words, attributes them to speakers, and assigns speaker timestamps, eliminating hand-offs between separate subsystems. The model is trained on extensive in-the-wild conversational audio—rich in accents, acoustic environments, overlaps, and domain shifts—to learn robust attribution and timing under realistic conditions. To preserve discourse and speaker coherence over long meetings, MOSS Transcribe Diarize is equipped with a \textbf{128k-token context window}, allowing it to process inputs up to 90 minutes without chunking. This long-context capability enables the model to build and maintain global representations of participants and topics, improving both diarization consistency and the handling of far-apart references.
Our contributions can be summarized as follows:
\begin{itemize}
    \item \textbf{End-to-end SATS system.} To our knowledge, MOSS Transcribe Diarize is the first unified multimodal model that jointly performs word recognition, speaker attribution, and timestamp prediction in a single forward pass.
    \item \textbf{128k-token long-context modeling at meeting scale.} MOSS Transcribe Diarize processes inputs up to \emph{90 minutes} within a \emph{128k}-token context window, preserving discourse continuity and long-range speaker memory without chunking; this reduces identity drift and boundary artifacts and improves SATS metrics across in-domain and out-of-domain evaluations.
\end{itemize}

\section{Model Architecture}

\begin{figure}[t!]
    \centering
    \includegraphics[width=0.9\textwidth]{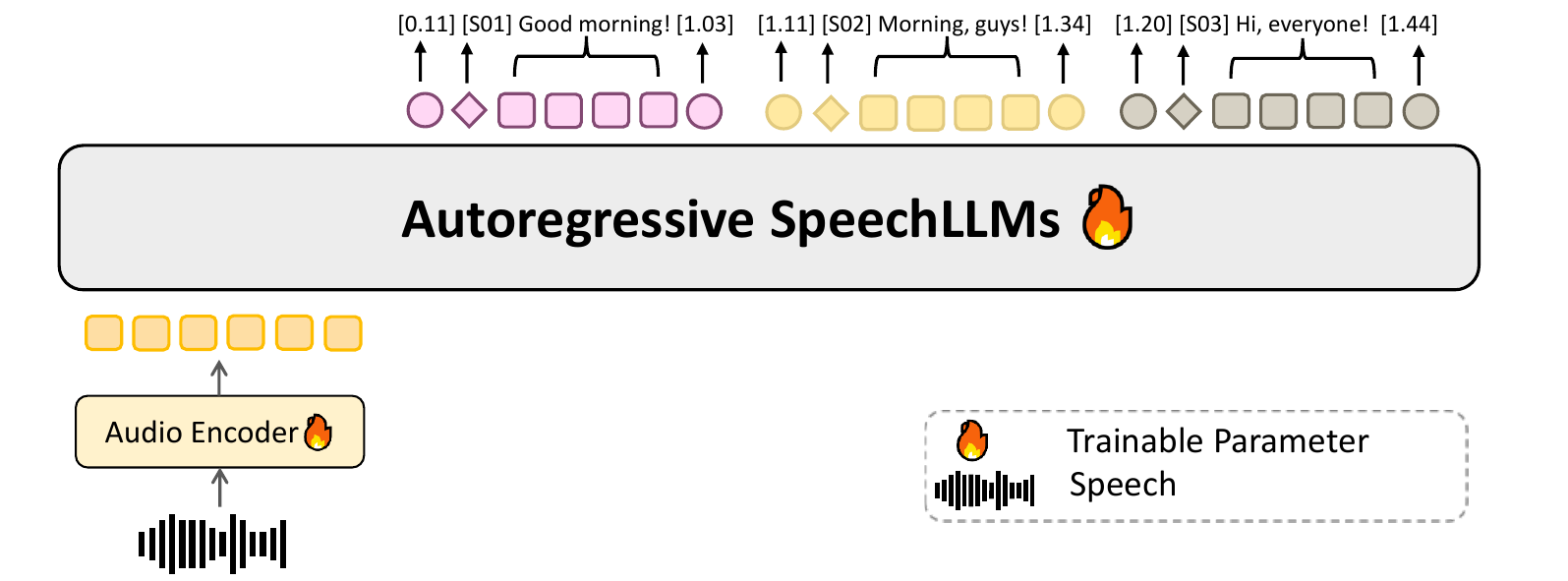}
    \caption{Overall architecture of the MOSS Transcribe Diarize model.}
    \label{fig:model_structure}
\end{figure}

\noindent 
As illustrated in Figure~\ref{fig:model_structure}, MOSS Transcribe Diarize couples an audio encoder with a projection module that maps multi-speaker acoustic embeddings into the feature space of a pretrained text LLM, enabling the backbone to jointly align speaker identities with lexical content and perform unified, long-context modeling in a single end-to-end model \citep{chu2023qwenaudio,tang2023salmonn,zhang2023speechgpt}.

Following recent work on textual token–based time encoding for long-context multimodal models \citep{chen2024timemarker,radford2023whisper}, we represent temporal information explicitly as formatted timestamp text inserted between audio encoder chunks. This avoids binding temporal encoding to absolute positional indices, which become sparse and ineffective over long durations, and enables accurate timestamp generation over hour-scale audio with stable speaker attribution \citep{bain2023whisperx}.

\section{Data Composition}
\subsection{Real Data}
In this study, we conduct experiments on multilingual audio collected from the Internet. We sample a large number of speaker-containing clips from public corpora for training, covering a wide range of real-world multi-speaker scenarios. In particular, the AISHELL-4 dataset \citep{Fu2021AISHELL4} comprises multi-speaker conversational recordings captured in meeting rooms, including both far-field overlapping audio and near-field recordings for each speaker. We use the averaged channel of the far-field signals for both training and evaluation. In addition, we further curated two datasets from podcasts and films to serve as test sets.

\subsection{Simulated Data}
To strengthen speaker attribution and timestamp prediction, and to cope with the scarcity of high-quality real-world recordings, we use simulated data during training. From our in-house corpus, we randomly sample a pool of single-speaker utterances to construct synthetic mixtures. Following previous work \citep{Park2023PropertyAware}, we employ a controllable probabilistic simulator to construct synthetic multi-speaker conversational data. Specifically, for each synthetic dialogue, we first draw 2–12 distinct speakers and randomly select one utterance per speaker. Each selected utterance is then partitioned into contiguous word runs by sampling a segment count and log-normal weights; the resulting segments are placed on a single timeline with Gaussian-distributed inter-segment gaps, enforcing speaker alternation while permitting overlaps capped at 80 percent of the shorter segment. To improve perceptual continuity, segment boundaries are snapped to nearby low-energy points and 50 ms cross-fades are applied. Following prior work \citep{Landini2022SimConv}, we augment the mixtures with real-world noise and reverberation, sampling SNRs uniformly from 0–15 dB.

\section{Evaluation}
\subsection{Evaluation Setups}

\subsubsection{Evaluation Datasets}

To comprehensively evaluate our model's performance, we use three diverse benchmarks. The \textbf{AISHELL-4 Test} set provides challenging, long-form audio from real-world conference scenarios. The \textbf{Podcast} set is composed of high-quality, multi-guest interviews from YouTube, using the platform's available subtitles which provide both reference transcripts. The \textbf{Movies} dataset consists of short audio segments derived from online films and TV series, which are rich in multi-speaker overlapping scenarios. It primarily features Chinese and English, but also covers other languages and dialects, including Korean, Japanese, and Cantonese. All samples in this dataset were manually annotated by professionals to ensure high-quality ground truth. The two internally curated datasets, Podcast and Movies, will be open-sourced and publicly released on Hugging Face to facilitate further research. The statistical overview of the datasets is provided in Table~\ref{tab:dataset_stats}.

\begin{table}[t!]
\centering
\begin{tabular}{l ccc}
\toprule
\textbf{Dataset} & \textbf{Duration Range (s)} & \textbf{Avg. Duration (s)} & \textbf{Number of Speakers} \\
\midrule
AISHELL-4 Test   & 2195.4 -- 2393.9          & 2290.6                   & 5 -- 7                  \\
Podcast   & 1528.7 -- 3636.5          & 2658.9                   & 2 -- 11                 \\
Movies & 0.418 -- 29.888 & 11.526 & 1 -- 6 \\
\bottomrule
\end{tabular}
\caption{Statistical overview of the evaluation datasets.}
\label{tab:dataset_stats}
\end{table}

\subsubsection{Metrics}
We adopt a comprehensive set of metrics to evaluate our system on both Automatic Speech Recognition (ASR) and Speaker Diarization (SD). Our primary metrics include Character Error Rate (CER), concatenated minimum-permutation CER (cpCER), and their difference, $\Delta\text{cp}$.

\textbf{CER} measures the performance of the ASR component by comparing the predicted transcript against the ground-truth text, without regard to speaker identities, using the standard minimum edit distance \citep{levenshtein1966}.

\textbf{cpCER} jointly evaluates both ASR and SD. It compares the predicted speaker-attributed transcripts against the ground-truth speaker transcripts. To handle label permutation ambiguity, the cpCER is calculated by finding the optimal assignment of predicted speaker labels that yields the minimum edit distance \citep{levenshtein1966,kanda2020sot,park2025sortformer}, thus reflecting the overall system performance for the speaker-attributed recognition task.

\textbf{$\Delta\text{cp}$} is the difference between cpCER and CER ($\Delta\text{cp} = \text{cpCER} - \text{CER}$). This value isolates the performance degradation caused by speaker attribution errors, thereby serving as a reliable, transcript-based measure of SD performance.

All metrics are reported in percentage (\%), where lower values indicate better performance. 

\begin{table*}[t!]
\centering
\renewcommand{\arraystretch}{1.3} 
\setlength{\tabcolsep}{4pt} 

\begin{threeparttable}
\small 
\begin{tabular}{ll ccccccc}
\toprule

\textbf{Dataset} & \textbf{Metric} & \textbf{Doubao} & \textbf{ElevenLabs} & \textbf{GPT-4o} & \textbf{Gemini 2.5 Pro} & \textbf{Gemini 3 Pro} & \textbf{VibeVoice} & \textbf{Ours} \\
\midrule
\multirow{3}{*}{\textbf{AISHELL-4}} 
& CER ($\downarrow$)   & 18.18 & 19.58 & \textbackslash & 42.70 & 22.75 & 21.40 & \textbf{14.84} \\
& cpCER ($\downarrow$) & 27.86 & 37.95 & \textbackslash & 53.42 & 27.43 & 24.99 & \textbf{15.83} \\
& $\Delta$cp ($\downarrow$) & 9.68  & 18.36 & \textbackslash & 10.72 & 4.68 & 3.59 & \textbf{0.99} \\

\midrule
\multirow{3}{*}{\textbf{Podcast}} 
& CER ($\downarrow$)   & 7.93  & 8.50  & \textbackslash & 7.38  & \textbackslash & 27.94 & \textbf{5.97} \\
& cpCER ($\downarrow$) & 10.54 & 11.34 & \textbackslash & 10.23 & \textbackslash & 48.30 & \textbf{7.37} \\
& $\Delta$cp ($\downarrow$) & 2.61  & 2.85  & \textbackslash & 2.85  & \textbackslash & 20.36 & \textbf{1.40} \\

\midrule
\multirow{3}{*}{\textbf{Movies}} 
& CER ($\downarrow$)   & 9.94  & 11.49 & 14.37 & 15.46 & 8.62 & 14.59 & \textbf{6.36} \\
& cpCER ($\downarrow$) & 30.88 & 17.85 & 23.67 & 24.15 & 14.73 & 42.54 & \textbf{12.76} \\
& $\Delta$cp ($\downarrow$) & 20.94 & 6.37 & 9.31  & 8.69  & \textbf{6.11} & 27.94 & 6.40 \\

\midrule
\multirow{3}{*}{\textbf{Alimeeting}} 
& CER ($\downarrow$)   & 25.25 & 25.70 & \textbackslash & 27.43 & 26.75 & 27.40 & \textbf{24.86} \\
& cpCER ($\downarrow$) & 37.57 & 36.69 & \textbackslash & 41.64 & 32.84 & 29.33 & \textbf{22.17} \\
& $\Delta$cp ($\downarrow$) & 12.31 & 10.99 & \textbackslash & 14.21 & 6.09 & 1.93 & \textbf{-2.69} \\

\bottomrule
\end{tabular}

\begin{tablenotes}
    \footnotesize
    \item \textit{Note:}  GPT-4o is omitted from the first two benchmarks due to its  audio input constraint. Gemini 3 Pro is excluded due to its instability in  adhering to the required output format for long audio inputs.
\end{tablenotes}
\end{threeparttable}
\caption{Performance of MOSS Transcribe Diarize and other models. Best results are in \textbf{bold}.}
\label{tab:comprehensive_results}
\end{table*}

\subsection{Performance}

To benchmark our model against current industry standards, we evaluate its performance relative to the best closed source models including Doubao Speech Recognition Model\footnote{\url{https://www.volcengine.com/docs/6561/1354871}}, ElevenLabs Scribe v1\footnote{\url{https://elevenlabs.io/docs/models\#scribe-v1}}, GPT-4o Transcribe Diarize\footnote{\url{https://platform.openai.com/docs/models/gpt-4o-transcribe-diarize}}, Gemini 2.5 Pro\footnote{\url{https://ai.google.dev/gemini-api/docs/models?hl=zh-cn\#gemini-2.5-pro}}, and Gemini 3 Pro\footnote{\url{https://ai.google.dev/gemini-api/docs/models?hl=zh-cn\#gemini-3-pro}}.
The results are shown in Table~\ref{tab:comprehensive_results}.

Table~\ref{tab:comprehensive_results} reports the performance of MOSS Transcribe Diarize in comparison with strong closed-source commercial systems across three representative benchmarks, covering long-form meeting recordings (AISHELL-4), extended multi-speaker conversations (Podcast), and short, overlap-rich segments (Movies). Together, these benchmarks evaluate robustness with respect to audio duration, speaker cardinality, and conversational structure.

Across all datasets where evaluation is feasible, \textbf{MOSS Transcribe Diarize consistently achieves the best overall performance} in terms of cpCER and $\Delta$cp, indicating superior joint modeling of transcription and speaker attribution. On AISHELL-4, which consists of nearly 40-minute real-world meeting recordings, our model substantially outperforms all baselines in both CER and cpCER. More importantly, it exhibits a markedly smaller $\Delta$cp, demonstrating that speaker attribution errors introduce significantly less additional degradation compared to pure ASR errors. This highlights the effectiveness of long-context, end-to-end modeling in maintaining speaker consistency over extended conversations.

It is worth noting that \textbf{GPT-4o and Gemini 3 Pro are unable to reliably process long-form audio inputs} such as AISHELL-4 and Podcast under our evaluation protocol. GPT-4o is constrained by its audio input length, preventing complete transcription of these recordings, while Gemini 3 Pro frequently fails to generate valid outputs that adhere to the required speaker-attributed format for long audio inputs. As a result, these systems are omitted from the corresponding benchmarks. This limitation underscores a practical gap between nominal multimodal capability and deployable long-form SATS performance.

On the Podcast benchmark, which features high-quality but long-duration, multi-speaker discussions, MOSS Transcribe Diarize again achieves the lowest CER and cpCER among all evaluated systems. While several baselines demonstrate strong ASR accuracy, our model consistently yields the smallest $\Delta$cp, indicating more reliable speaker attribution under frequent turn-taking and long-range speaker re-entrance. This advantage is particularly important for real-world conversational analytics, where speaker identity coherence across long temporal spans is critical.

The Movies dataset presents a complementary challenge characterized by short utterances, rapid speaker alternation, and frequent overlaps. Even in this short-form setting, MOSS Transcribe Diarize outperforms all baselines in cpCER and $\Delta$cp. Notably, some commercial systems achieve competitive CER but suffer from substantially larger $\Delta$cp values, reflecting difficulties in resolving speaker attribution under dense overlap. In contrast, our model maintains a relatively small gap between CER and cpCER, indicating robust handling of speaker boundaries across diverse conversational regimes.

Overall, these results demonstrate that the advantages of MOSS Transcribe Diarize extend beyond improved word recognition accuracy. The consistently low $\Delta$cp across both long-form and short-form benchmarks confirms that the proposed end-to-end SATS formulation, together with long-context modeling, yields more reliable speaker-attributed, time-stamped transcripts than modular or semi-cascaded alternatives. Crucially, unlike several general-purpose multimodal models, MOSS Transcribe Diarize remains fully operational on hour-scale audio, making it particularly well suited for real-world meeting transcription and long-form conversational analysis.

\section{Conclusions}
We introduced \textbf{MOSS Transcribe Diarize}, a unified audio--text MLLM for Speaker-Attributed, Time-Stamped Transcription (SATS) that jointly performs transcription, speaker attribution, and timestamp prediction in a single pass with a \textbf{128k-token} context window. The model combines a speech encoder with a learned projection into a pretrained text LLM, and is trained on diverse in-the-wild conversations together with property-aware simulated mixtures that model overlap, turn-taking, and acoustic variability. Across AISHELL-4, Podcast, and Movies, MOSS Transcribe Diarize outperforms strong closed-source systems in CER, cpCER, and $\Delta\text{cp}$, highlighting the effectiveness of long-context joint modeling and distribution-controlled simulation for end-to-end SATS at meeting scale. Future work includes streaming SATS, finer-grained timestamp evaluation, and broader multilingual robustness.

\section*{Contributors}

\textbf{Core Contributors} \\
Donghua Yu$^{*}$, Zhengyuan Lin$^{*}$, Hanfu Chen$^{*}$, Chen Yang, Yiyang Zhang, Zhaoye Fei

\vspace{0.8em}

\textbf{Contributors} \\
Botian Jiang, Jingqi Chen, Ke Chen, Qinyuan Cheng, Liwei Fan, 
Yitian Gong, Yi Jiang, Muchen Li, Shimin Li, 
Songlin Wang, Wenxuan Wang, Yang Wang, Zhiyu Wu, 
Zhe Xu, Wenbo Zhang, Yuqian Zhang, Jie Zhu

\vspace{0.8em}

\textbf{Advisors} \\
Xipeng Qiu$^{\dagger}$

\vspace{1.0em}

{\small
$^{*}$ Equal contribution. \\
$^{\dagger}$ Corresponding author.
}

% ===== Bibliography =====
\clearpage
\bibliographystyle{plainnat}
\bibliography{main}

% ===== Appendix (content unchanged) =====
\clearpage
\beginappendix

\startcontents[app]
\begingroup
  \renewcommand{\contentsname}{Appendix Contents}
  \section*{\contentsname}
  \printcontents[app]{}{1}{}
\endgroup
\newpage

\section{Additional Details}
\subsection{Evaluation Prompts}

\newtcolorbox{mossbox}[1]{
    enhanced,                  
    colback=white,             
    colframe=MossBlue,         
    colbacktitle=MossBlue,    
    coltitle=black,           
    title style={left color=MossCyan, right color=MossBlue},
    title=\textbf{#1},         
    fontupper=\ttfamily,               
    boxrule=0.8pt,      
    left=2mm, right=2mm, top=2mm, bottom=2mm 
}

\begin{mossbox}{Gemini 2.5 Pro \& Gemini 3 Pro (AISHELL-4, Podcast, Movies)}
<audio>\\
请将音频中的对话内容转换为文本，语言使用音频的语言，使用[S1],[S2]...等标签标注出所有说话人。\\
比如：[S1]你好[S2]你好[S1]你叫什么名字？[S2]我叫小明。\\
只需要输出最终的结果，不需要输出其他任何内容，不需要输出换行符\\ 
\end{mossbox}

\begin{mossbox}{MOSS Transcribe Diarize Prompts}
\textbf{AISHELL-4 \& Podcast} \\
<audio>\\
请将音频转写为文本，每一段需以起始时间戳和说话人编号（[S01]、[S02]、[S03]…）开头，正文为对应的语音内容，并在段末标注结束时间戳，以清晰标明该段语音范围。

\vspace{0.3cm}
\hrule
\vspace{0.3cm}

\textbf{Movies} \\
<audio>\\
请将以下对话转录为文本，使用 [S1] [S2] 等说话人标签，对于音频中的事件，使用 [event] 标签表示。富有情感的文本用<emotion>对应文本</emotion> 表示，使用 <ovl> 标签表示音频有部分重叠，<ins></ins> 标签表示音频有插入。自动检测音频的语言，说话人标签和 <ovl> <ins> 始终用英文，event 和 emotion 跟随音频语言。
\end{mossbox}

\subsection{Output Normalization for Evaluation}
To ensure fair comparison across systems, we apply the same text normalization to both predictions and references before computing CER/cpCER/$\Delta$cp.
Given a raw string $x$, we perform the following steps:

\begin{itemize}[leftmargin=1.5em]
    \item \textbf{Remove parenthetical content.}
    We delete any text in parentheses (and the preceding whitespaces) using
    \texttt{\textbackslash s*\textbackslash(.*?\textbackslash)}.
    \item \textbf{Remove angle-bracket tags.}
    We delete any substrings matching \texttt{<.*?>} (e.g., \texttt{<emotion>...</emotion>}, \texttt{<ovl>}, \texttt{<ins>...</ins>}).
    \item \textbf{Remove non-speaker square-bracket annotations.}
    We delete any square-bracketed spans \texttt{[...]} that are not speaker IDs, using the regex
    \texttt{\textbackslash[(?!S\textbackslash d+\textbackslash]).*?\textbackslash]}.
    This keeps only speaker tags of the form \texttt{[S1]}, \texttt{[S01]}, etc., and removes other bracketed markers such as events (e.g., \texttt{[event]}).
\end{itemize}

After normalization, each hypothesis/reference contains only speaker identifiers and plain transcript text for scoring.

\end{document}